# One-step stirring preparation of room temperature liquid metal negative electrode for the lithium-ion battery

Yao Huang[a], Yibin Jiang[a], Haijuan Wang[a], Xunyong Jiang[a],*

*e-mail: jiangxunyong@tjut.edu.cn

[a] School of Materials Science and Engineering, Tianjin University of Technology, Tianjin, China

**Abstract**: As a new type of self-healing material, room-temperature liquid metal (LM) composed of Ga, In and Sn is a promising anode in lithium-ion batteries (LIBs). It is difficult to directly prepare an electrode slurry with pure LM. Here, the LM electrode slurry was successfully prepared by simple high-speed stirring method. The LM in the slurry was evenly distributed in the form of liquid particles (average size 100 μm) and no aggregation occurred. The initial discharge specific capacity of LM is 1148 mAh $g^{-1}$, and the stable discharge specific capacity is 350 mAh $g^{-1}$. Cyclic voltammetry measurements show that three pairs of obvious reduction peaks and oxidation occur in LM anodes at 0.37 V, 0.67 V, 1.02 V and 0.72 V, 0.79 V, 0.97 V. In the process of lithium insertion, the LM undergoes a liquid-solid phase transition, and when it is delithiated, it will be converted into a liquid phase. The SEI film consumed active lithium and electrolyte during the long cycle, resulting in loss of reversible capacity.

**Key words:** Ga-In-Sn, Liquid metal, Lithium Ion Battery, negative electrode,

# 1 Introduction

Due to the limitation of fossil energy and environmental pollution, it becomes more urgent to develop an energy storage device with high energy density, high stability and long cycle life[1-4]. Among them, lithium-ion batteries are widely used in commercial, portable equipment, automotive and other fields and have good application prospects. Therefore, it is of great significance to develop a negative electrode material with excellent lithium storage performance[5].

At present, the widely used commercial graphite anodes have a lower actual capacity(372 mAh g$^{-1}$), which greatly limits the improvement of lithium ion battery performance[6]. It is extremely urgent to find new anode materials to meet the growing demand[5,7]. Alloy-based anode materials, such as silicon and tin, have become one of the key materials for next-generation high-energy-density ion batteries due to their high theoretical capacity and low delithiation potential. However, the large volume change of the alloy anode material during the charge and discharge process will cause the alloy particles to crack and smash, and even fall off from substrate, which will cause the battery performance to decline. Suppressing the volume change of the alloy anode material is benefit for improving its cycle property[6-14].

Ga is a low-melting-point metal with a melting point of 29.8 ℃. The theoretical capacity is 769 mAh g$^{-1}$[14-18]. Zhang[19] found that at 40 ℃, the metallic Ga in the liquid state has a high discharge specific capacity (518 mAh g$^{-1}$) and good charge-discharge cycle performance.Ga form Li$_2$Ga alloy when it is fully intercalated with lithium, and becomes a solid state. After delithiation, it becomes a liquid state.

Ga undergoes a solid-liquid phase change during the deintercalation of lithium, which can repair surface cracks caused by deintercalation of lithium[17,20-22]. Liquid Ga electrode have self-healing property during charge/discharge cycle.

Ga-based alloys doped with appropriate amounts of In and Sn have lower melting points than pure Ga[23-26]. It is capable of maintaining liquid state below room temperature (25℃)[27]. This greatly increases the range of application of liquid metal anode materials. For example, 1) Chen and others used it as a lithium-ion anode, and charge and discharge tests showed that it has good cycling performance[15]. 2) Chen et al[5] synthesized a type of self-healing core-shell fibers, with LM nanoparticles as the core coated with a carbon shell by facile coaxial electrospinning and a carbonization process, which fix the liquid metal in the carbon nanotubes. The test results show that it has a high discharge specific capacity and good cycle stability. 3) Seung et al[28] prepared a liquid gallium electrode material by vaporization and deposition on a porous carbon basis, and fixed the liquid metal by using porous carbon. 4) Jinkui Feng synthesize LM flexible electrode material using Mxene as carrier[24]. The procedure used in these papers to prepared liquid metal (LM) electrode is as follow: 1) Adding LM into some liquid with surfactant in it; 2) Ultrasonic crushing is then performed to obtain stable LM droplets; 3) Finally, other substances such as Mxene are added to fix small droplets of LM for subsequent electrode slurry preparation.The whole process is not simple enough, and it is difficult to use in large-scale production. Moreover, the nanomaterial used in these papers to fix LM droplets have certain lithium storage capacity itself. This will affect the analysis of the capacity and the lithium storage mechanism of the LM itself. What is the capacity of room temperature LM? What are the structural changes of LM during charge and discharge? The answer of these question depend on the preparation of pure room temperature LM electrode. Unfortunately, the Ga-based alloy in the liquid state at room temperature is prone to agglomeration has brought great challenges to the electrode material pulping.

High-speed stirring is widely used in preparing electrode slurry for LIBs. As far as we know, there is no report on the direct preparation of LM electrodes by high-speed stirring up to now. Can agitation be used to break up liquid metal to form small droplets? We first tried to add LM directly to the right amount of N-methyl-2-pyrrolidone (NMP) for high-speed stirring. It was found that the LM can be shredded with stirring, but aggregation occurs after the stirring is stopped. In order to avoid agglomeration of LM during stiring, LM was mixed with conductive agent in an appropriate amount of NMP for high-speed stirring. The result shows that the LM was effectively broken into small particles. The LM is uniformly distributed in the electrode slurry composed of the conductive agent and the binder. LM does not agglomerate. When the slurry is coated on a Cu substrate and dried, the LM is stably present on the electrodes, LM is well bonded to the Cu current collector tightly.

Here, we successfully prepared pure LM electrode by simple high-speed stirring method without using additional additives. The LM is evenly distributed on the current collector with no agglomeration. LM electrode material shows a high discharge specific capacity and good cycle stability. The stable discharge specific capacity is 350 mAh g$^{-1}$. The mechanism of lithium storage in LM and the morphological changes in the process of lithium intercalation were analyzed. In the process of lithium insertion, the LM undergoes a liquid-solid phase transition, and when it is delithiated, it will be converted into a liquid phase.

## 2 Materials and method

The LM used in this article is a commercial Ga-In-Sn alloy (Ga:62 wt%; In:25 wt%; Sn:13 wt%) with a melting point of 5 ℃. The LM electrodes for electrochemical test were prepared as follow: the LM and PVDF were mixed at a mass ratio of 8:1. The mixture with addition of a suitable amount of Nmethyl-2-pyrrolidone (NMP) were stirred with a magnetic whisk for 1 h (2000 r min$^{-1}$), and pasted onto the copper foils (coating thickness 100 μm). Dry the materials at 25 ℃ for 1h and then vacuum dry at 120 ℃ for 12h. Finally punch them into circle electrode sheets (16 mm in diameter). A porous polyethylene film (Celgard 2400) was used as a separator, metal lithium was used as a positive electrode, 1M LiBF$_4$ dissolved in EC, DEC and EMC mixture of equal volumes was used as electrolyte. The coin cells were assembled in an Ar-filled glove box where moistuer and oxygen contents were strictly controlled to below 1 ppm.

The samples were characterized using Rigaku D/Max-2500 V X-ray diffraction (XRD) in the range of 10°-90°. Charge-discharge cycles test of the electrodes were performed at 25 ℃ using a NEWARE tester. Electrochemical impedance spectroscopy (EIS) was tested using an IM6 electrochemical workstation in a frequency range of 110 Hz-0.01 Hz with AC amplitude 5 mV. Cyclic voltammetry measurements were tested at a scan rate of 0.1 mV s$^{-1}$ in the voltage range of 0-3 V and current range of -40 to 40 mA. Analyses of surface composition of the LM electrodes before and after cycling were proformed using XPS (Kratos-AXIS UL TRADLD, Al KaX-ray source). The morphology and elemental changes of the LM electrode sheets before and after cycling were analyzed using field emission scanning electron microscope (FESEM, JSM-6700 F).

## 3 Results and discussion

### 3.1 Preparation of LM electrode slurry

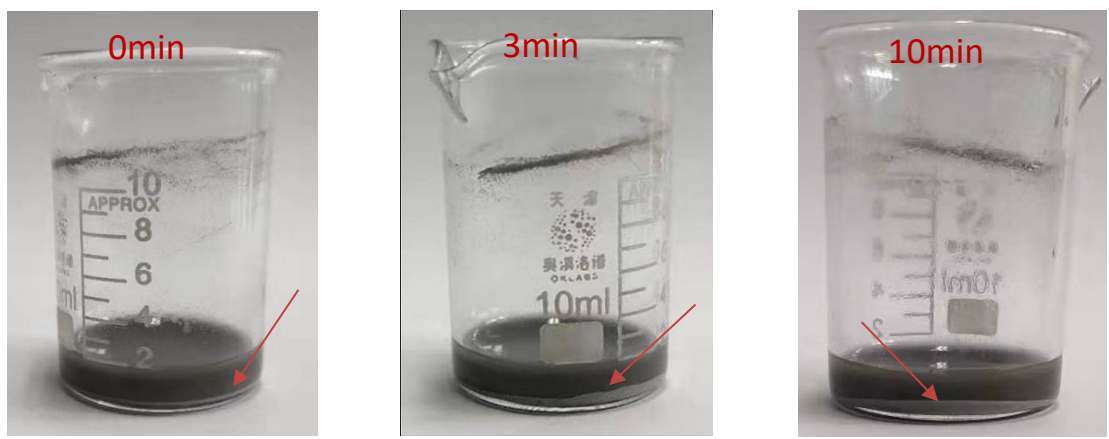

Figure 1 Slurry state at different times after stopping stirring

High-speed stirring is widely used in preparing electrode slurry for LIBs. We first mixed LM and NMP directly by high speed stirring (2000 r min$^{-1}$). The result is shown in Figure.1. It can be seen that a uniform suspension is obtained after stirring which indicates that high-speed stirring can break the LM into small drops (Figure.1a,b). However, LM drops stick together quickly after stopping the agitation. Bright LM forms a sediment layer in the lower part of the solution (Figure.1c).

Acetylene black is often used as a conductive agent in the preparation of electrode slurry for lithium ion batteries. It has a large surface area with strong adsorption capacity. LM drops may absorbed on the surface of acetylene black which may anchor LM drops to avoid its agglomeration in the NMP solution. LM, PVDF, and acetylene black were mixed at a mass ratio of 8:1:1, in an appropriate amount of NMP. The solution is stirred with high speed (2000 r min$^{-1}$) for 1h. After the stirring, the obtained slurry was uniform (Figure.2a). No agglomeration occurred (Figure.2b). The slurry can be evenly coated on the copper substrate. The surface of the coated electrode sheet was flat after drying. The active substance binds well with the substrate.It can be seen that some white bright spots are evenly distributed in the active substance(Figure.2c). These bright spots are LM drops. From the SEM image, The size of drops is about 100 μm (Figure.2d). The above results show that mixing LM with acetylene black in NMP by using high-speed stirring can effectively break the LM into small droplets. The acquired LM droplets are stable in the active material slurry. The electrode were successfully prepared with this slurry for the subsequent lithium storage property test.

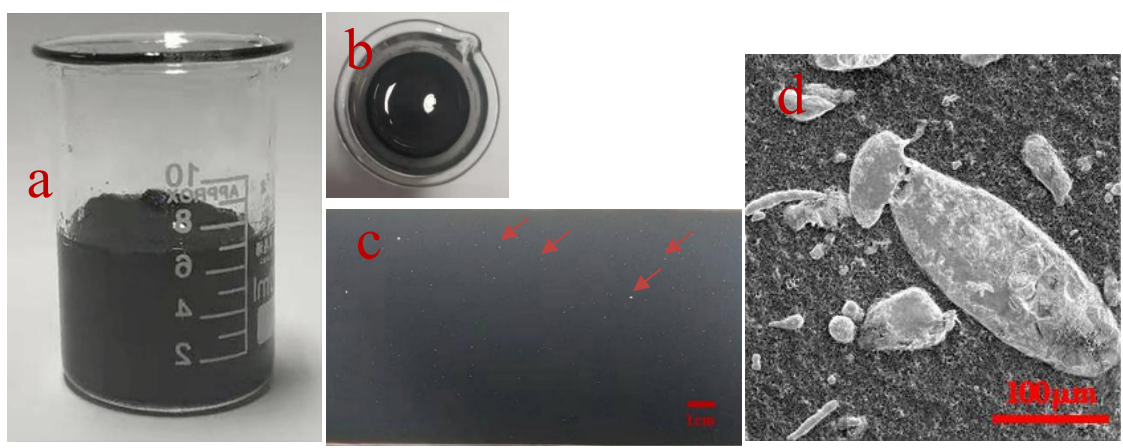

Figure 2 a b is the obtained slurry, c is the electrode sheet,d is the SEM image of the electrode sheet

3.2 Electrochemical performance of LM

The electrochemical performance of pure room temperature LM electrodes was explored by galvanostatic discharge/charge measurements, and CV. The results are shown in Figure.3. The first cycle discharge specific capacity of the LM electrode with a current density of 60.5 mA $g^{-1}$ was 1148 mAh $g^{-1}$. After 16 cycles, the discharge specific capacity was stable at 350 mAh $g^{-1}$ (as shown in Figure.3a). Compared with the first cycle, the reversible capacity retention rate of LM is only 30%. But in the subsequent high-current charge and discharge tests of 121 mA $g^{-1}$, 242 mAh $g^{-1}$, 363 mAh $g^{-1}$, 484 mAh $g^{-1}$, and 605 mAh $g^{-1}$, it has shown good cycle stability, high coulomb efficiency (well above 90%), high reversible capacity retention rate (as shown in Table 1). After 75 cycles of high-rate, when the current returned to 60.5 mAh $g^{-1}$, the discharge specific capacity suddenly increased and stabilized at 275 mAh $g^{-1}$. The LM anode material has excellent charge-discharge performance and high reversible capacity retention rate, especially under high rate charge and discharge conditions. The LM anode material still has good charge and discharge performance after a large current and long cycle.

From the charge and discharge curve (Figure 3b), it can be seen that when the current density is 60.5 mA $g^{-1}$, there is a small potential plateau at 1.3 V during the first discharge, which corresponds to the formation of a SEI film on the electrode surface[28]. The formation of the SEI film resulted in a reversible capacity loss in the first cycle (Figure 3a). There are three potential platforms for the LM electrode during the first discharge, which correspond to the inclined platforms of 1.2 V, 0.8 V, and 0.5 V, respectively. These three potential platforms are related to the lithium ion insertion process. Phase transition reactions of Ga to $Li_2Ga_7$, $Li_2Ga_7$ to LiGa, and LiGa to $Li_2Ga$ occur respectively[22,26]. Lithium ion extraction corresponds to 0.5 V, 0.7 V, and 0.9 V inclined platforms during charging. Phase changes of $Li_2Ga$ to LiGa, LiGa to $Li_2Ga_7$, and $Li_2Ga_7$ to Ga respectively occur[20,22,26]. As the cycle progresses, the SEI film will grow and become thicker and will stabilize. At the same time, the insertion and extraction of lithium ions becomes more difficult. So all platforms disappear completely in the 2nd and 16th cycles.

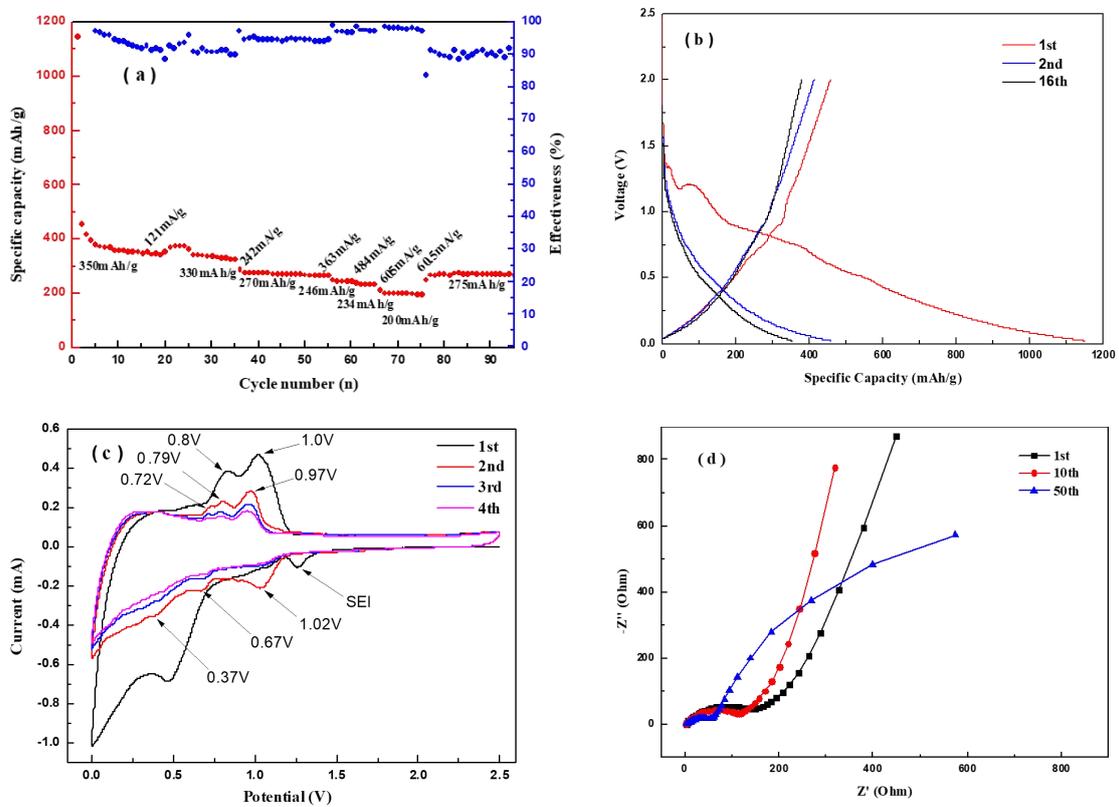

Figure 3 Electrochemical test of Ga-In-Sn (a) cycling performance at different currents (b) charge-discharge curves. (c) CV curves. (d) Nyquist plots of Ga-In-Sn at an open-circuit potential with different cycle.

Table 1 Reversible capacity retention rate of charge and discharge at different current densities

| Current density | 121 mA g$^{-1}$ | 242 mA g$^{-1}$ | 363 mA g$^{-1}$ | 484 mA g$^{-1}$ | 605 mA g$^{-1}$ | 60.5 mA g$^{-1}$ |
|---|---|---|---|---|---|---|
| Reversible capacity retention rate (%) | 94 | 77 | 70 | 67 | 57 | 78.6 |

The above results are consistent with the CV test. From the results of cyclic voltammetry (Figure.3c), it can be seen that a broad reduction peak appears at 1.25 V in the first cycle, which is related to the SEI film formed on the electrode surface during the first charge and discharge. The broad reduction peak disappeared in the subsequent cycles, indicating that the SEI film formation tended to stabilize as the charge-discharge cycle progressed. In the first cycle, two distinct redox peaks appeared at 1.0 V and 0.8 V, 0.5 V and near 0 V, respectively. In the following second, third and fourth cycles, three distinct pairs of redox peaks appeared at 0.72 V, 0.79 V, 0.97 V and 0.37 V, 0.67 V and 1.02 V, respectively. This is the same as the three inclined platforms in Figure.3b, which correspond to different phase change reactions. This shows that during the cycle, Li$^+$ can be normally inserted and released in LM. As the charge-discharge cycle progresses, the currents of the oxidation peak and the reduction peak gradually decrease. It shows that the ability to deintercalate lithium decreases, and the polarization of the electrode sheet increases. Furthermore, the potential difference of the redox reaction is gradually increased, and the reversibility of the battery is reduced. This is consistent with the test results of constant current charge and discharge. In addition, the positions and shapes of the peaks in the second, third and fourth cycles are basically the same, indicating that the material has relatively high chemical reversibility, which also explains the excellent cycle stability of the LM anode material at higher current density . Figure.3d is the impedance spectrum of the LM electrode. The semicircle in the high frequency region is due to the formation of the SEI film and the process of lithium ions passing through the SEI film. The oblique line in the low frequency region corresponds to the diffusion process of lithium ions in the LM electrode material. It can be seen that the semicircle of the high frequency region of the electrode material of the first period is the largest, and the slope of the linear region of the low frequency region is the smallest. This is because a large number of the deintercalated lithium channels are not completely opened in the first cycle, thus hindering the diffusion of lithium ions. As the charge and discharge cycle progresses, the channel gradually opens and the impedance decreases.

3.3 Structural change of LM during charge-discharge cycle

In order to study the phase evolution of LM during charge and discharge cycle, we performed XRD tests on LM in different states. The results are shown in Figure.4(a). Curve (1) shows acetylene black as the conductive agent. Curve (2), (3), and (4) are the LM electrode in the fully charged, fully discharged, and original state, respectively. Since the thickness of the electrode coating is only a few tens of micrometers, a diffraction peak of Cu appears on the electrode sheet substrate.

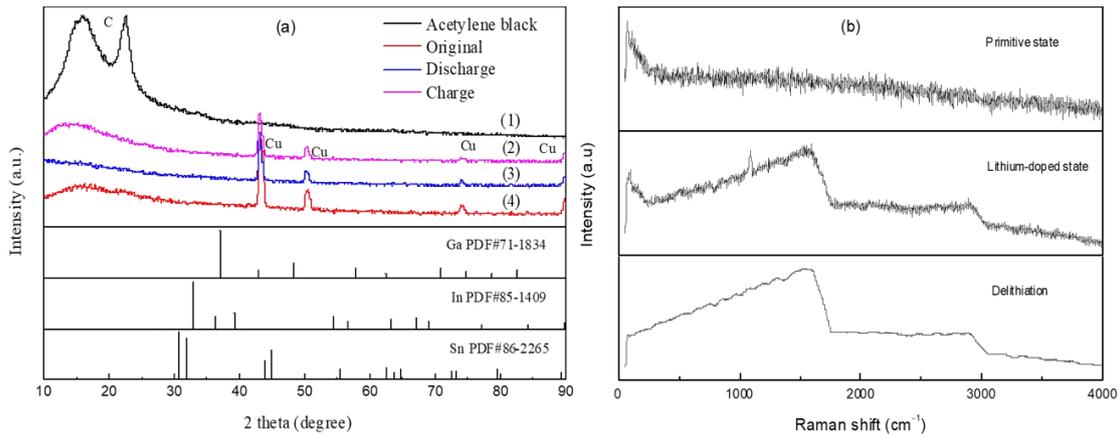

Figure 4 (a)XRD patterns of Ga-In-Sn liquid metal electrode sheet. Curve (1) is the test result of pure acetylene black, and curve (2) (3) (4) is the fully charged, fully discharged, unreacted electrode piece test result.(b) Raman testing of liquid metals in different states

The X-ray diffraction peaks of the three metal elements Ga, In and Sn and their alloy compounds were not found in the figure. In addition, the X-ray diffraction peaks of the sample before and after charging were not significantly changed. It shows that no new crystalline phase is formed during the charge and discharge cycle. It can be seen from Figure.4(b) that the original LM surface is in a completely amorphous state. After lithium insertion, three distinct peaks appeared at 1000 $cm^{-1}$, 1600 $cm^{-1}$, and 2900 $cm^{-1}$. This indicates that a crystalline phase was formed on the LM surface. This is because the lithium intercalation becomes solid during the LM discharge. After delithiation, the peak at 1000 $cm^{-1}$ disappeared completely and the curve became smooth. This is because the LM metal is delithiated and becomes liquid, and the surface becomes amorphous. However, the peaks at 1600 $cm^{-1}$ and 2900 $cm^{-1}$ did not change significantly, and unknown crystal phases may be formed.

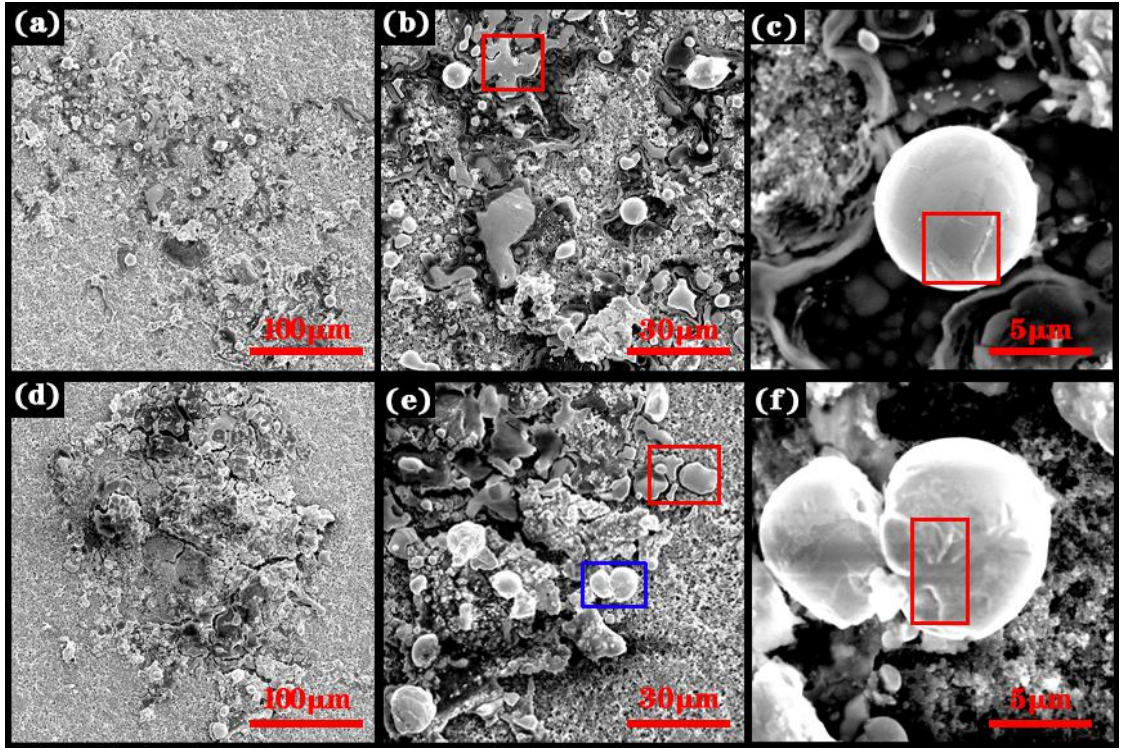

Figure 5 SEM images of Ga,In,Sn on electrodes at (a),(b),(c) fully discharged, (d),(e),(f) fully recharged states

Before the reaction, most of the LM particles are completely coated with acetylene black (Figure.2d). A small part of the exposed LM is rod-shaped, relatively close to the elliptical shape, and the surface is rough and uneven, which is due to the stress of the high-speed agitation process. it is beneficial for the electrolyte to fully contact with the active material to facilitate the insertion of lithium ions, which can also explain the high discharge specific capacity in the first cycle of the charge-discharge test. The morphology of the LM electrode at different states were characterized by SEM, as shown in Figure.5. After the complete discharge, most of the LM is exposed on the surface of the electrode. The LM rod becomes a small with size of only about 10 μm. Some of the particles even are spherical (Figure.5 (a),(b)). The insertion of lithium ions causes the LM to change from a liquid to a solid. The volume expands, which causes the surface of LM cracks (The red selection in

Figure.5(c))[27], and the large particles break into small particles. It is can be seen from the red region of Figure.5(e) that some membranous substances are gradually formed between the electrode surface and the LM particles, which makes some particles not present in a sphere. This film-like substance corresponds to the formation of the SEI film. The SEI film of the electrode surface grows thicker as the charge-discharge cycle proceeds. It has an increasingly obvious effect on the morphology of LM particles. It can be seen from Figure.5(e)(f) that during the charging process, the particle size of the LM particles does not further decrease, but a slight increase tends to occur (Figure.5(f)). There are two main reasons for this: on the one hand, part of the lithium insertion is difficult to escape during charging and discharging, so that the LM cannot change from solid to liquid[8, 11]; on the other hand, when the LM particles are small to a certain extent, the LM after delithiation can repair the micro cracks on the surface to some extent (red selection in Figure.5(f)) without cracking[19,21]. At the same time, due to the continuous growth of the SEI film (the red selection in Figure.5(e)) and the surface tension, the original independent LM particles begin to be connected to each other, and agglomeration occurs, as shown in Figure.5(e) (the red selection). In addition, the formation of the SEI film on the electrode surface also consumes the electrolyte and lithium in the battery, which also explains the large reversible capacity loss that occurs in the charge-discharge test.

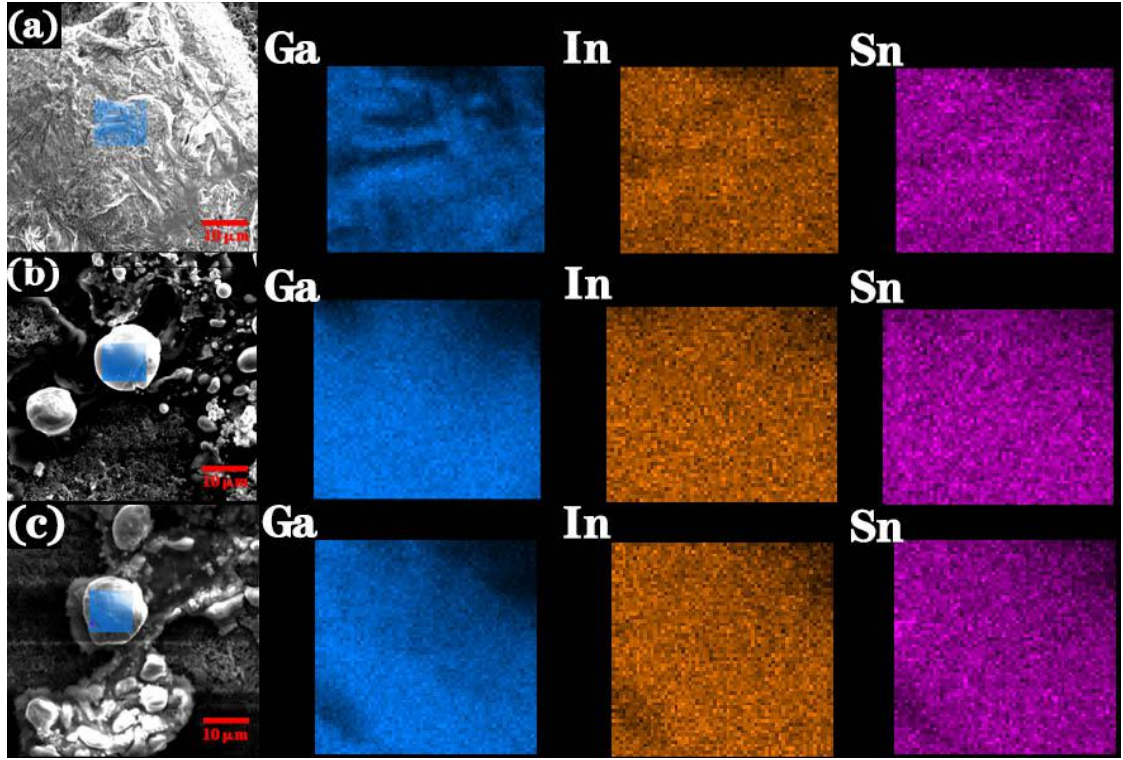

Figure 6 SEM images and SEM-EDS mapping of Ga,In,Sn elemental distributions on electrodes at (a) pristine, (b) fully discharged, and (c) fully recharged states.

The distribution of elements in the LM at different state,before the reaction, fully discharged and fully charged state,were shown in Figure.6. It can be seen that the three elements of Ga, In and Sn on the LM surface are evenly distributed, indicating that the LM still exists in the form of alloy during the charge-discharge process, and there is no precipitation or agglomeration of three elements of Ga, In, and Sn.

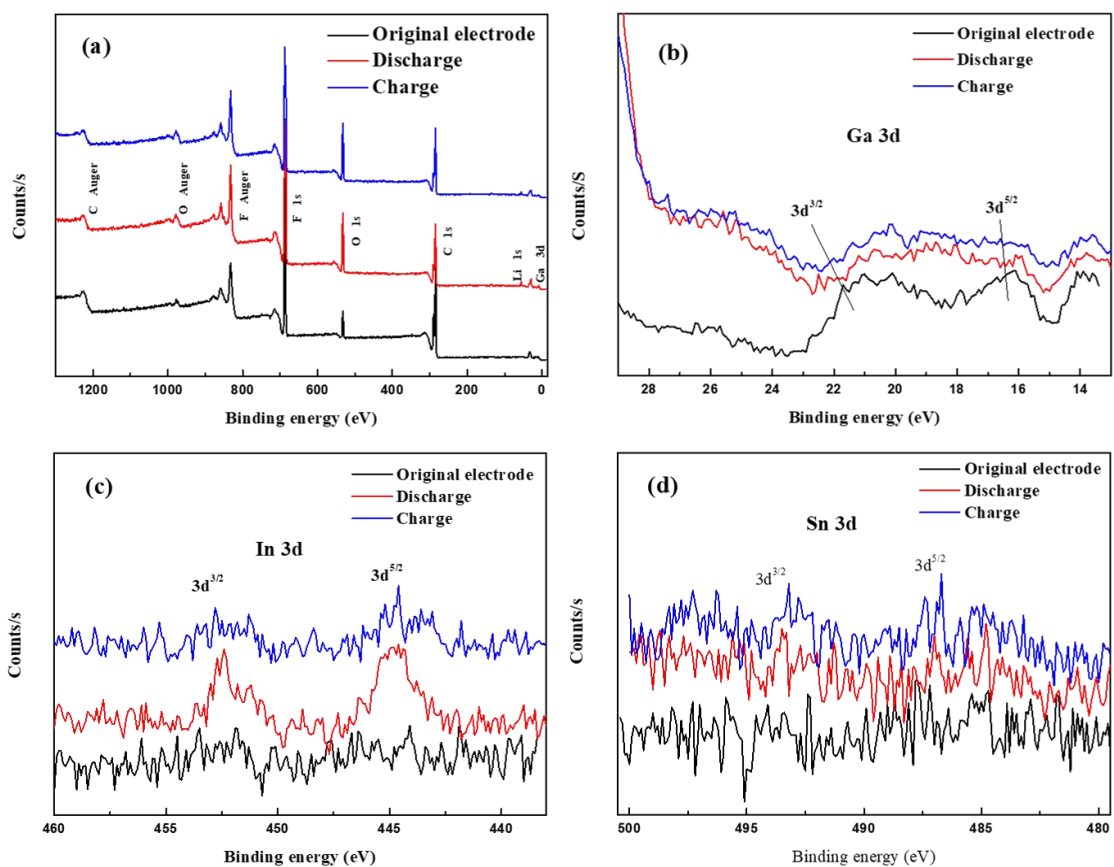

Figure 7 (a) XPS survey spectra, (b) Ga 3d spectra, (c) In 3d, (d) Sn 3d

XPS test were performed on the LM negative electrode before and after charging and discharging to characterize and analyze the changes of elements and compounds during the process of deintercalating lithium. A clear peak of Ga 3d can be seen from Figure.7(a). However, since the content of Sn and In in the active material is very low, no significant peak appears. Compared with the original state, the Li 1s peak appears on the full spectrum after deintercalation of lithium, which is mainly caused by the growth of the SEI film during the charge-discharge cycle and the difficulty in the partial release of lithium. As shown in Figure.7(b), after the charge-discharge cycle, the peak of Ga 3d shifts slightly to the right, and the binding energy decreases, possibly generating $GaF_3$. The peak and binding energy changes of In 3d and Sn 3d in

Figure.7(c)(d) are not particularly obvious, indicating that the LM discharge specific capacity is mainly derived from Ga. In, Sn mainly plays a role in regulating the melting point of LM to 5 ℃ to ensure that the LM electrode can work normally at room temperature.

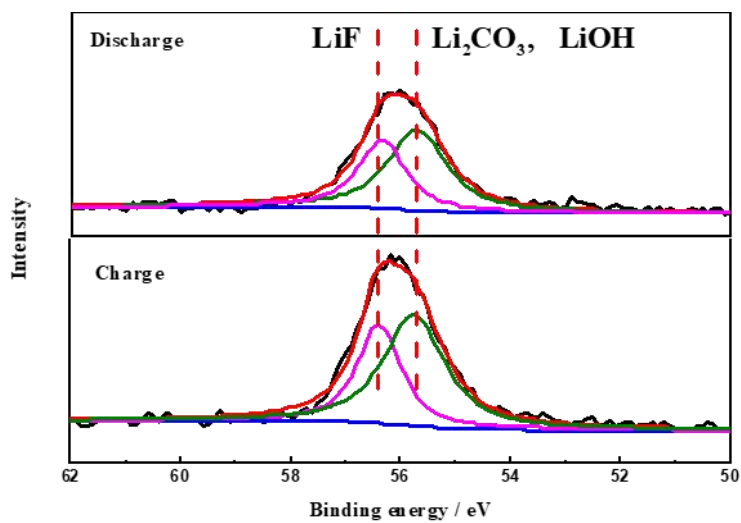

Figure 8 Li1s XPS spectra

The Li 1s peak are shown before and after charge and discharge in Figure.8. It can be clearly seen from the figure that LiF, $Li_2CO_3$ and LiOH are formed on the surface of the LM during charging and discharging. These compounds are the main components of the SEI film, which is consistent with the SEM test results. As the cycle progress, the SEI film continues to grow, and the SEI film is continuously thickened and its composition changes may reduce the migration ability of lithium ions in the SEI film. At the same time, the continuous growth of the surface SEI film reduces the amount of active lithium in the battery, resulting in a decrease in the reversible capacity of the battery. This also explains the large reversible capacity decay that occurs during the previous cycle.

## 4. Conclusions

In summary, We have used the industrially established magnetic stirring method for the scalable preparation of room temperature LM electrode. A series of electrochemical tests and characterizations were carried:

1) Using a high-speed stirring method with a conductive agent and a binder, the LM is successfully broken into uniformly distributed liquid particles. The acquired LM droplets are stable in the active material slurry. The slurry can be evenly coated on the copper substrate. The active substance binds well with the substrate.

2) The LM material has good charge and discharge cycle performance, and there is an irreversible process during the first charge and discharge process, and the discharge specific capacity is reduced to 350 mAh g$^{-1}$ after 16 cycles.

3) The LM is still in a liquid state after charging and discharging cycle. It has self-repairing properties.

4) Two pairs of obvious redox peaks appear in the cyclic voltammetry test chart, indicating that the LM electrode material has better deintercalation lithium performance.

5) The growth of the SEI film on the surface of the electrode during the charging and discharging process consumes active lithium, and also causes the liquid metal particles to agglomerate, resulting in loss of reversible capacity.

## 5 Prospect

In this paper, a LM metal electrode material was successfully prepared by high-speed stirring method. The method has the advantages of simple operation, low cost and low process requirement. Moreover, it can be used for large-scale electrode material preparation. In the future, the minimum size of the liquid metal LM can be reduced by adjusting the stirring conditions, thus the charge and discharge performance and cycle stability of the LM electrode material can be improved.

# References:

[1]S. Chen, J. Wang, F. Ling, R. Ma, B. Lu, ADV ENERGY MATER, 8(2018) 1800140.

[2]M.C. Lin, M. Gong, B. Lu, Y. Wu, H. Dai, Science Foundation in China, v.23(2015) 21.

[3]A.S. Arico, P. Bruce, B. Scrosati, J.M. Tarascon, W. van Schalkwijk, NAT MATER, 4(2005) 366-377.

[4]S. Fang, D. Bresser, S. Passerini, ADV ENERGY MATER, (2019).

[5]J. Zhu, Y. Wu, X. Huang, L. Huang, M. Cao, G. Song, X. Guo, X. Sui, R. Ren, J. Chen, NANO ENERGY, 62(2019) 883-889.

[6]W. Zhang, J POWER SOURCES, 196(2011) 13-24.

[7]Y. Shi, M. Song, Y. Zhang, C. Zhang, H. Gao, J. Niu, W. Ma, J. Qin, Z. Zhang, J POWER SOURCES, 437(2019) 226889.

[8]W. Tu, Z. Bai, Z. Deng, H. Zhang, H. Tang, NANOMATERIALS-BASEL, 9(2019).

[9]Y. Zhang, Y. Zhu, L. Fu, J. Meng, N. Yu, J. Wang, Y. Wu, CHINESE J CHEM, 35(2017) 21-29.

[10]T. Kasukabe, H. Nishihara, S. Iwamura, T. Kyotani, J POWER SOURCES, 319(2016) 99-103.

[11]B. Wang, J. Yang, J. Xie, K. Wang, Z. Wen, X. Yu, ACTA CHIM SINICA, 61(2003) 1572-1576.

[12]S. Iwamura, H. Nishihara, Y. Ono, H. Morito, H. Yamane, H. Nara, T. Osaka, T. Kyotani, SCI REP-UK, 5(2015).

[13]W. Zhang, J POWER SOURCES, 196(2011) 877-885.

[14]T. Sakai, Y. Xia, T. Fujieda, K. Tatsumi, M. Wada, H. Yoshinaga, Insertion Electrode Materials for Rechargeable Lithium Batteries, Studies in Surface Science and Catalysis, Elsevier, 2001, pp.939-942.

[15]Y. Wu, L. Huang, X. Huang, X. Guo, D. Liu, D. Zheng, X. Zhang, R. Ren, D. Qu, J. Chen, ENERG ENVIRON SCI, 10(2017) 1854-1861.

[16]Y. Wu, X. Huang, L. Huang, X. Guo, R. Ren, D. Liu, D. Qu, J. Chen, ACS Applied Energy Materials, 1(2018) 1395-1399.

[17]M.W. Verbrugge, R.D. Deshpande, J. Li, Y. Cheng, MRS Proceedings, 1333(2011).

[18]Q. Zhanga, JingLiua, NANO ENERGY, (2013) 863-872.

[19]张素林，罗飞，郑杰允，褚赓，刘柏男，化学学报，73(2015) 808-814.

[20]F. Yu, J. Xu, H. Li, Z. Wang, L. Sun, T. Deng, P. Tao, Q. Liang, Progress in Natural Science: Materials International, 28(2018) 28-33.

[21]S. Tang, D.R.G. Mitchell, Q. Zhao, D. Yuan, G. Yun, Y. Zhang, R. Qiao, Y. Lin, M.D. Dickey, W. Li, Matter, 1(2019) 192-204.

[22]R.D. Deshpande, J. Li, Y. Cheng, M.W. Verbrugge, J ELECTROCHEM SOC, 158(2011) A845.

[23]X. Guo, L. Zhang, Y. Ding, J.B. Goodenough, G. Yu, ENERG ENVIRON SCI, 12(2019) 265-2619.

[24]C. Wei, H. Fei, Y. Tian, Y. An, G. Zeng, J. Feng, Y. Qian, SMALL, (2019) 1903214.

[25]Q. Yu, Q. Zhang, J. Zong, S. Liu, X. Wang, X. Wang, H. Zheng, Q. Cao, D. Zhang, J. Jiang, APPL SURF SCI, 492(2019) 143-149.

[26]S. Chen, L. Wang, Q. Zhang, J. Liu, SCI BULL, 63(2018) 1513-1520.

[27]W. Liang, L. Hong, H. Yang, F. Fan, Y. Liu, H. Li, J. Li, J.Y. Huang, L.Q. Chen, T. Zhu, S. Zhang, NANO LETT, 13(2013) 5212-5217.

[28]K.T. Lee, Y.S. Jung, T. Kim, C.H. Kim, J.H. Kim, J.Y. Kwon, S.M. Oh, Electrochemical and Solid-State Letters, 11(2008) A21.